\documentclass[twocolumn,aps,prl]{revtex4}
\usepackage{graphicx}
\usepackage{amsmath}
\usepackage{amssymb}
\usepackage{wasysym}
\usepackage{color}

\newcommand{\unit}[1]{\ensuremath{\, \mathrm{#1}}}

\begin{document}


\title{How micropatterns and air pressure affect splashing on surfaces}

\author{Peichun Tsai, Roeland van der Veen, Matthias van de Raa, and Detlef Lohse}

\affiliation {Physics of Fluids Group, Faculty of Science and Technology,
Impact \& MESA$^{+}$ Institutes, J. M. Burgers Center, University of Twente, 7500AE
Enschede, The Netherlands.}
\date{\today}

\begin{abstract}
We experimentally investigate the splashing mechanism of a millimeter-sized ethanol drop impinging on a structured solid surface, comprised of micro-pillars, through side-view and top-view high speed imaging. By increasing the impact velocity we can tune the impact outcome from a gentle deposition to a violent splash, at which tiny droplets are emitted as the liquid sheet spreads laterally. We measure the splashing threshold for different micropatterns and find that the arrangement of the pillars significantly affects the splashing outcome. In particular, directional splashing in direction in which air flow through pattern is possible. Our top-view observations of impact dynamics reveal that an trapped air is responsible for the splashing. Indeed by lowering the pressure of the surrounding air we show that we can suppress the splashing in the explored parameter regime.
\end{abstract}


\maketitle

\section{Introduction}
A high-speed, wetting drop that impacts onto a solid surface can generate a splash, emitting small secondary droplets from a spreading lamella at the impact.
The complex interplay between the droplet type, surface property, and surrounding gas not only produces surprising outcomes but also obscures the underlying mechanism~\cite{Yarin_review,MRein_ActaMech2008,modeling_splashing}.  Recent studies have shown previously un-foreseen effects in splashing impact onto a solid surface; for example, a splash can be eliminated by reducing gas pressure~\cite{LXu_PRL_2005}, by increasing the height of the pillars forming textured substrates~\cite{LXu_PRE_2007_2}, by decreasing the tension of an impacted elastic membrane~\cite{HStone_PoF_2008}, by controlling relative tangential velocity between a droplet and a dry, smooth surface~\cite{HStone_NJPhy_2009}. In addition, a recent theory reasons the entrapment of a thin gas film as a precursor to the splash onto a solid, smooth surface~\cite{MBrenner_PRL_2009}.  The predicted gas film is very thin, in the order of magnitude O($0.1 - 1~\mu m$), with a fast formation dynamics in O($0.1~\mu s$)~\cite{MBrenner_JFM_2010}. As a result, the detection of such gas film poses a great experimental challenge. Yet, to our best knowledge, no direct observations of the liquid-gas interface upon a solid surface prior to splashing exist in the literature.

\begin{figure}
\begin{center}
\includegraphics[width=3in]{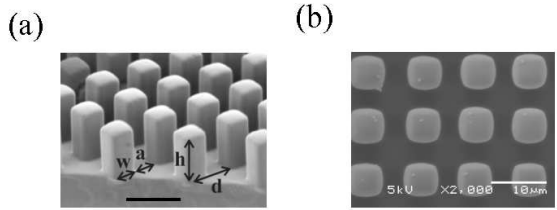}
\vspace{-0.1in}
\caption{\label{microstructures}
(a) Scanning electron microscope (SEM) image of a representative micropatterned surface used in the ethanol drop impact experiment; (b) shows the top-view of the microstructures. Both inset bars have the same length scale of $10 \mu m$.}
\end{center}
\end{figure}

\begin{figure}[h]
\begin{center}
\includegraphics[width=3.4in]{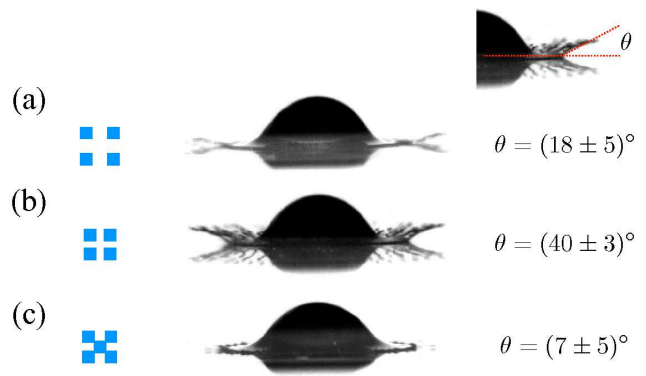}
\caption{\label{side_view_1atm}
(Color online; see the supplementary movies) Side-view snapshots reveal distinctive ethanol splashing dynamics at $0.4$ ms on different microtextures, schematically shown by the top-view drawings of the used micro-molds, with two squared lattices of the periodicity of approximately $d = 10 \mu m$ and $d = 7 \mu m$ in (a) and (b), respectively, as well as with one chessboard-liked pattern in (c). $\theta$ is the tilting angle of the ejecting thin sheet and varies with the micro-patterns. The experimental conditions are $We = 380 \pm 10$, $R = 1.3$ mm, and $P_{air} = 1$ atm.}
\end{center}
\end{figure}

To fill in this gap, in this study we exploit {\it transparent} surfaces comprised of micro-pillars (see Fig.~\ref{microstructures}) to examine the role of the air film in the splashing impact onto rough surfaces. In addition, motivated by the crucial role of the initial contact of the droplet and the surface suggested by~\cite{Thoroddsen_JFM_2005, MBrenner_PRL_2009, MBrenner_JFM_2010}, we study microscopic wetting dynamics of the impacting droplets on a variety of rough substrates.  From our experience, the liquid-gas interface is rather difficult to observe from the high-speed top-view recordings of drop impact onto smooth surface.  As a result, here we only focus on the drop impact onto rough surfaces. Experimentally, a millimeter-sized ethanol droplet was impinged onto the surfaces protruded with micro-pillars in different microscopic arrangements. A striking observation was that slightly varying micro-patterns can result in distinct splashing dynamics, as illustrated in Fig.~\ref{side_view_1atm}: the angle of the ejecting liquid sheet and the intensity of splashing were altered with different micro-textured surfaces.  To gain further insight into the splashing mechanism we also recorded high-resolution, top-view snapshots of the wetting dynamics, in addition to controlling the surrounding air pressure. The average spatial resolution is about $30~\unit{\mu m/pixel}$ (with the highest resolution at $5~\unit{\mu m/pixel}$) recorded at a rate between $10~000$ and $30~000$~fps (frame per second). The top-view images (Fig.~\ref{top_view_1atm}, \ref{top_views_lowP}) reveal the areas wetted by the liquid.  Depending on the impact kinetics and the interspace between the micro-pillars, the central impact zone can be completely wetted by the drop or largely dry due to substantial air entrapment. Whether or not a large amount of air is entrapped results in different splashing dynamics. Under reduced air pressure, the dry, air-trapping central impact zone observed at $1$ atm turns into a largely wetted area and the splash is eliminated. 

\section{Experimental Section}
Our experimental setup and procedure are similar to those in Ref.~\cite{PTsai_langmuir_2009}.  Instead of water droplets used in those studies, here we focus on wetting-drop impact using ethanol (Merck chemicals, purity $\geq 99.9\%$, liquid density $\rho = 789 \unit{kg/m^{3}}$, surface tension $\sigma = 22.3 \times 10^{-3}$ Nm$^{-1}$, and viscosity $\mu = 1.2\times 10^{-3}$ kg/ms).  An ethanol droplet was released from a fine needle ($0.34$~mm, $0.51$~mm, or $1.07$ mm inner diameter), with a syringe pump (PhD 2000 infusion, Harvard Apparatus), at different heights to vary the impact velocity onto the microstructured surfaces. The substrate and needle were enclosed in a chamber connected to a vacuum pump so as to control the pressure of the surrounding air, $P_{air}$.  The substrate material is transparent PDMS (Polydimethylsiloxane, SYLGARD 186 Silicone Elastomer), which facilitates the top-view observations of wet and dry areas at the impact.  The fabrication of the microstructures is achieved with a micro-molding method by etching the inverse, desired microstructures on a silicon wafer as a master replica mold, which was cleaned with Pirana cleaning (a mixture of sulfuric acid $H_{2}SO_{4}$ and hydrogen peroxide $H_{2}O_{2}$, $5:1$ in volume ratio, for $30$ mins) and subsequently with ultrasonic cleaning in an ethanol bath. The clean mold then was hydrophobized by a vapor deposition of an alkylsilane (1H,1H,2H,2H-perfluorodecyltrichlorosilane) to allow for an easy release of the sample from the micropatterned mold. A de-gassed mixture of the PDMS base and the curing agent ($10:1$ mass ratio) was casted on the hydrophobized mold and subsequently cured in an oven at $85^\circ$C for three hours. Finally, the PDMS sample was peeled off from the wafer and used as the targeted surface. For the data presented here we employed a new sample for each experiment.

\begin{figure}
\begin{center}
\includegraphics[width=3.4in]{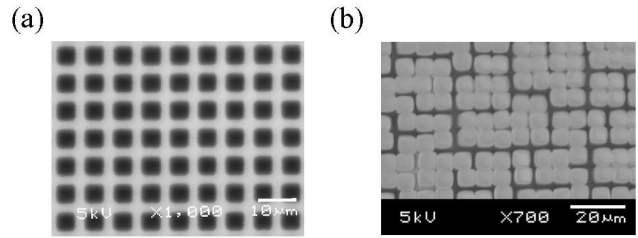}
\caption{\label{mold_vs_sample}
SEM images of (a) a representative replica mold of regular micro-holes in a square arrangement of the periodicity $d = 7.07~\mu m$ and (b) the produced sample, showing a maze-like structure formed by the bundled micro-pillars of the interspace $a \lesssim 2~\mu m$.}
\end{center}
\end{figure}

To investigate the effect of micro-arrangments, we used $6$ different molds of micro-patterns: cylindrical or rectangular micro-holes of the same depth $h = 6~\mu m$ and width $w = 5~\mu m$, arranged in a square, hexagonal, or chessboard-like lattice.  We moreover also varied the closest interspace $a$ between micro-pillars or the periodicity of the lattice unit $d$.  It is worthy to note that a precise micro-fabrication was achieved for a larger interspace $a = 5~\mu m$ (see Fig.~\ref{microstructures}), whereas bundles of tilting micro-pillars could occur for a smaller interspace $a \lesssim 2~\mu m$.  Fig.~\ref{mold_vs_sample} shows the latter example: the left image (a) shows a fraction of the  replica mold with which the sample was fabricated; (b) is the SEM image of the resulting product, illustrating a maze-like arrangement owing to the touching and tilting of relatively ``tall" pillars ($h = 6 \mu m$, and $a \approx 2~\mu m$) with the neighbors. Intriguingly, a similar observation was reported on the nanoscale: assemblies of bundled nano-fibers were found in the quest for a synthesis of an array of straight nano-fibers in a regular nano-arrangement~\cite{Geim_NatMat_2003}. 
In spite of the bundled structures, here the closely packed micro-pillars present a paradigmatic case by offering high impact resistance for the impinging droplet, modeling surfaces with roughness of $\mu m$. We note that the corresponding large solid packing fraction $\Phi_s$ remains unaffected by the bundling; $\Phi_s$ is the solid areal fraction of the pillars: $\Phi_s = \pi w^2/4d^2$ for cylindrical pillars in a squared lattice.   
Each sample has an areal dimension of $\approx 22.5 \times 22.5~\unit{mm}^2$ fully decorated with the micro-pillars. In presenting the data, we denote the geometric parameters of the micro-patterns based on the replica molds in the figures. In addition, in the texts we describe the exact dimensions analyzed with the SEM images.  We found that the chessboard-like patterns actually have wider pillars than the design of the replica mold, which may result from the limitation of the etching technique for producing the cross-linking micropatterns. 

The Weber number $We = \rho R V^2/\sigma$, as the main control parameter, compares the kinetic energy to surface energy. Here, the liquid density is denoted $\rho$, the drop radius $R$, the impact velocity $V$, and the surface tension $\sigma$.  Our $We$-numbers are between $100$ and $500$, limited by the height of the pressure-controlled chamber. The impact velocity $V$ ranges from $2.1$ to $4.1$ $ms^{-1}$. The Reynolds number gives the ratio of the inertia to viscous forces of the liquid drop: $Re = \rho RV/\mu$, with a large value ranging from $1050$ to $2500$. The capillary number compares the viscous force to surface tension: $Ca = V\mu/\sigma = We/Re$; our $Ca$-number lies in the range of $0.11-0.22$. The impact velocity $V$ and droplet diameter, $2R$, and their errors were directly analyzed from the recorded snapshots using our customer-made matlab programs, using a linear fit of traveling distance vs. time (usually from $10$ frames prior to the impact moment) and an elliptic-profile fitting of the droplet. 

\begin{figure*}
\begin{center}
\includegraphics[width=5in]{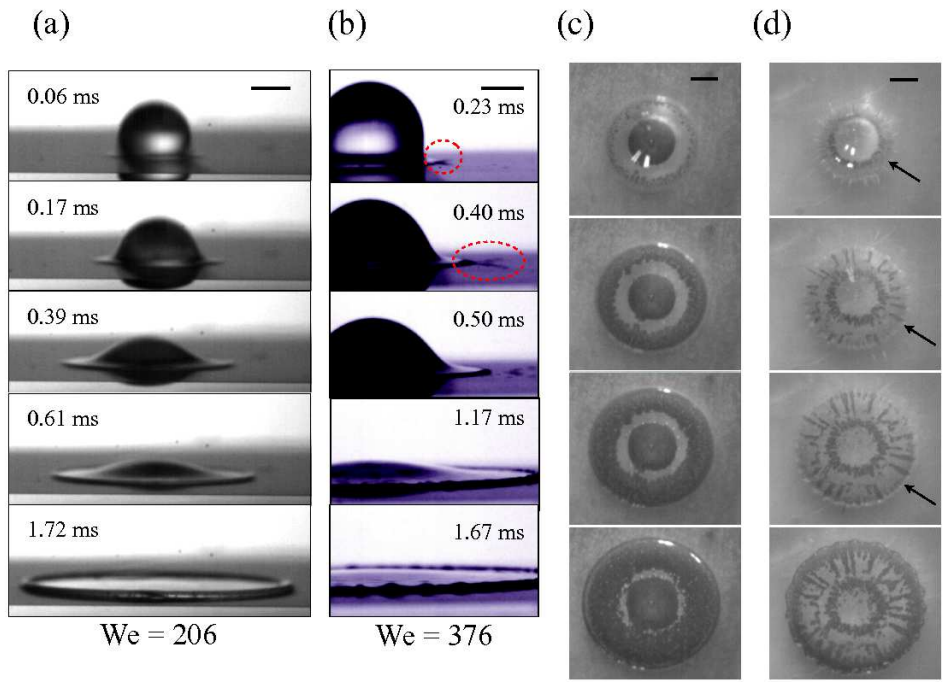}
\caption{\label{impact_events}
The time evolutions reveal two typical phenomena observed in the experiments of an ethanol drop impacting on micro-patterned substrates: (a) and (c) gentle deposition, i.e., simple spreading of a liquid sheet, and (b) and (d) violent splashing, i.e., emitting of small droplets within $0.5$ ms during the advancing phase of a spreading lamella. The inset bars indicate $1~\unit{mm}$ in length. The effect of Weber number is revealed by the side-views of spreading (a) and of splashing (b) for the same micro-patterned molds consisting of cylindrical holes of $w = 5~\mu m$ in width and $h = 6~\mu m$ in height in square arrangements, at different Weber numbers. The influence of micropatterns is shown by the top-view high speeding recordings at similar Weber number for square pillars with different interspaces $a = 5~\mu m$ in (c) spreading at $We = 189$ and $a \approx 2~\mu m$ in (d) splashing at $We = 184$. The time sequences in both (c) and (d) are the same, at $t = 0.25, 0.45, 0.65,$ and $1.05~$ms. The dark areas in (c) and (d) are wetted regions by ethanol. The arrows in (d) mark the edge of the lamella.
}
\end{center}
\end{figure*}
   
\section{Results and Discussion}

Two common scenarios; deposition and splash, as illustrated in Fig.~\ref{impact_events}, were observed in our studied parameter regime with ethanol. A deposition shows a simple spreading of liquid sheet with no discharge of small droplets.  A splash displays an emission of tiny secondary droplets, jetting out through the rim of the spreading lamella. We characterize a "splashing" event for the outcome with at least one secondary droplet discharged. A splash can be generated simply by increasing the Weber number as revealed by Fig.~\ref{impact_events} (a) and (b) for the same kind of micropatterned substrates. Fig.~\ref{impact_events} (c) and (d) also reveals a profound effect of the targeted micro-patterns on the impact outcome.  For a similar $We$-range of $O(200)$, spreading occurs for micro-pillars of the larger interspace $a = 5~\mu m$ in (c) while splashing for a smaller interspace $a \approx 2~\mu m$ in (d).  The splash occurs in an early time stage: slightly upward jetting droplets were apparent within $0.5$~ms after the impact. This type of splash should be distinguished from the formation of satellite droplets during the receding phase of the lamella; for instance, the so-called "receding breakup" and "partial rebounding" events~\cite{Yarin_review, MRein_ActaMech2008}, which were not observed in the experiments due to the wetting liquid. In comparison, for water drop impact on similar surfaces decorated with hydrophobic micro-pillars different events, such as bouncing, jetting, entrapping of bubble, and a partial bouncing, were found for $We$ of O(1-10)~\cite{DQuere_Nature_2002, DQuere_Nature_2003, DQuere_bouncing_transition_EPL2006, bartolo_jets_PRL2006, PTsai_langmuir_2009}.

\begin{figure}
\begin{center}
\includegraphics[width=3.4in]{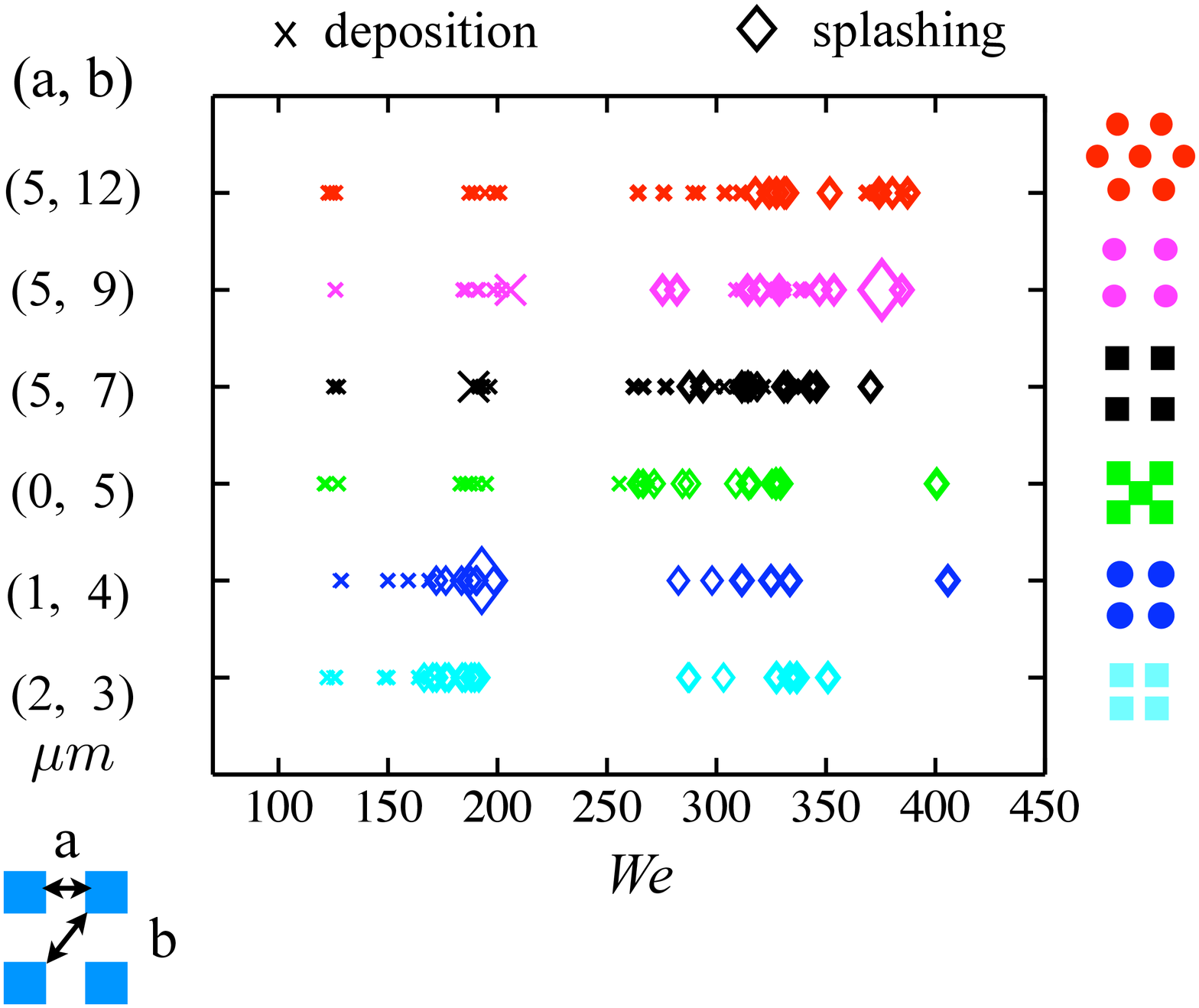}
\caption{\label{phase_diagram}
(Color online) A fraction of the phase space of the ethanol droplet impact dynamics at $P_{air} = 101.3$ kPa on the microstructured surfaces, showing the influences of the micro-patterns and the Weber number on the impacting behavior, either depositing ($\times$) or splashing ($\diamond$). Different colors indicate a variety of the used micro-molds, as schematically shown by the top-views of the unit pattern of the replica mold and expressed in terms of geometric parameters: the closest interspace $a$ and the largest interspace $b$ in a single pattern unit.  The time evolutions shown in Fig.~\ref{impact_events} correspond to the data points with the symbols printed in large.}
\end{center}
\end{figure}

\subsection{The influences of microstructures and the Weber number}
\begin{figure}
\begin{center}
\includegraphics[width=3.4in]{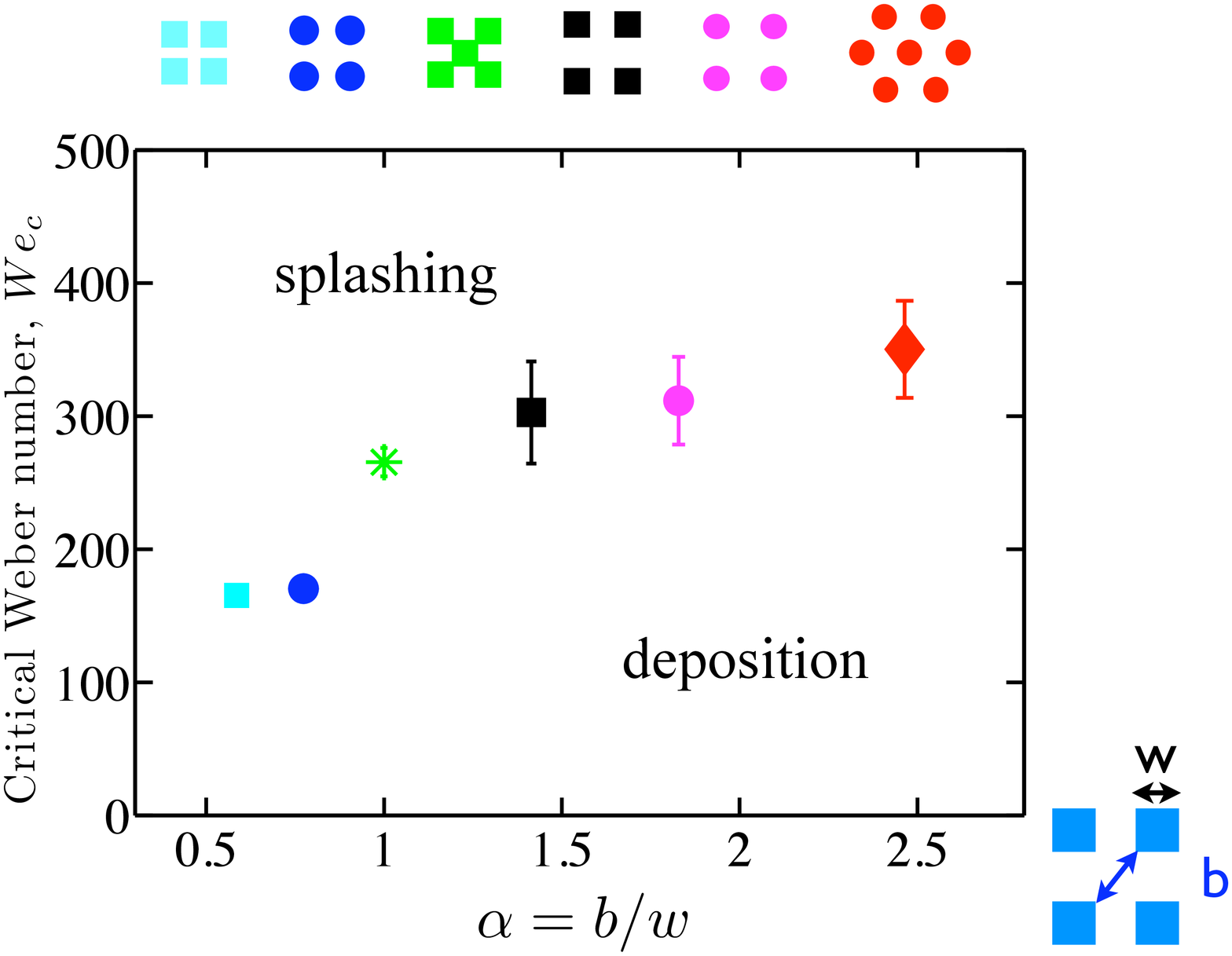}
\caption{\label{critical_We} (Color online) The influence of micropatterns on the critical Weber number above (below) which a splashing (deposition) takes place. The dimensionless length parameter, $\alpha$, is defined as the ratios $b/w$ of the length scales of the replica mold. Here the width of the pillars $w$ is fixed, and $\alpha$ describes the ratio of the largest length scale for air passing to the width of the obstacles in one unit of the micro-pattern.
}
\end{center}
\end{figure}

Fig.~\ref{phase_diagram} is the phase diagram of ethanol drop impact at $P_{air} = 101.3~\unit{kPa}$ onto the microtextured surfaces, showing the effects of the micro-pattern and the impact velocity. In general, droplets with small impact velocity deposit on all the micro-patterns for $100 \lesssim We \lesssim 150$, whereas with larger $We$ droplets splash. 
We mark the threshold of Weber number identifying the boundary between deposition and splash for each micro-pattern.  Fig.~\ref{critical_We} shows the critical Weber number $We_c$ (or the so-called splashing threshold) above (below) which a splash (deposition) occurs for a particular microstructure. $We_c$ increases for a larger $\alpha$, which is a length ratio of the largest space in a unit micro-pattern $b$ to the pillar width $w$ (see the sketch in Fig.~\ref{phase_diagram}). The micropatterns of $\alpha < 1$, with $We_c \approx 167$, are maze-liked and closely-packed and thus provide high resistance for the impacting droplet to reach the bottom surface as well as for the intervening air to drain out. In contrast, a stronger impact is needed for splashing onto the regularly arranged micro-pillars with a wider interspace ($\alpha > 1$). The error bars in $We_c$ indicate the overlapping regimes when both splashing and deposition are observed in the experiments. Here changing the interspace of the micro-patterns $a$ by $\approx 3~\mu m$ can result in more than $50\%$ increase in the critical Weber number $We_c$. Different shapes (squared or round) of the micro-pillars in similar squared lattice arrangements have negligible influence on $We_c$.

\begin{figure*}
\begin{center}
\includegraphics[width=6in]{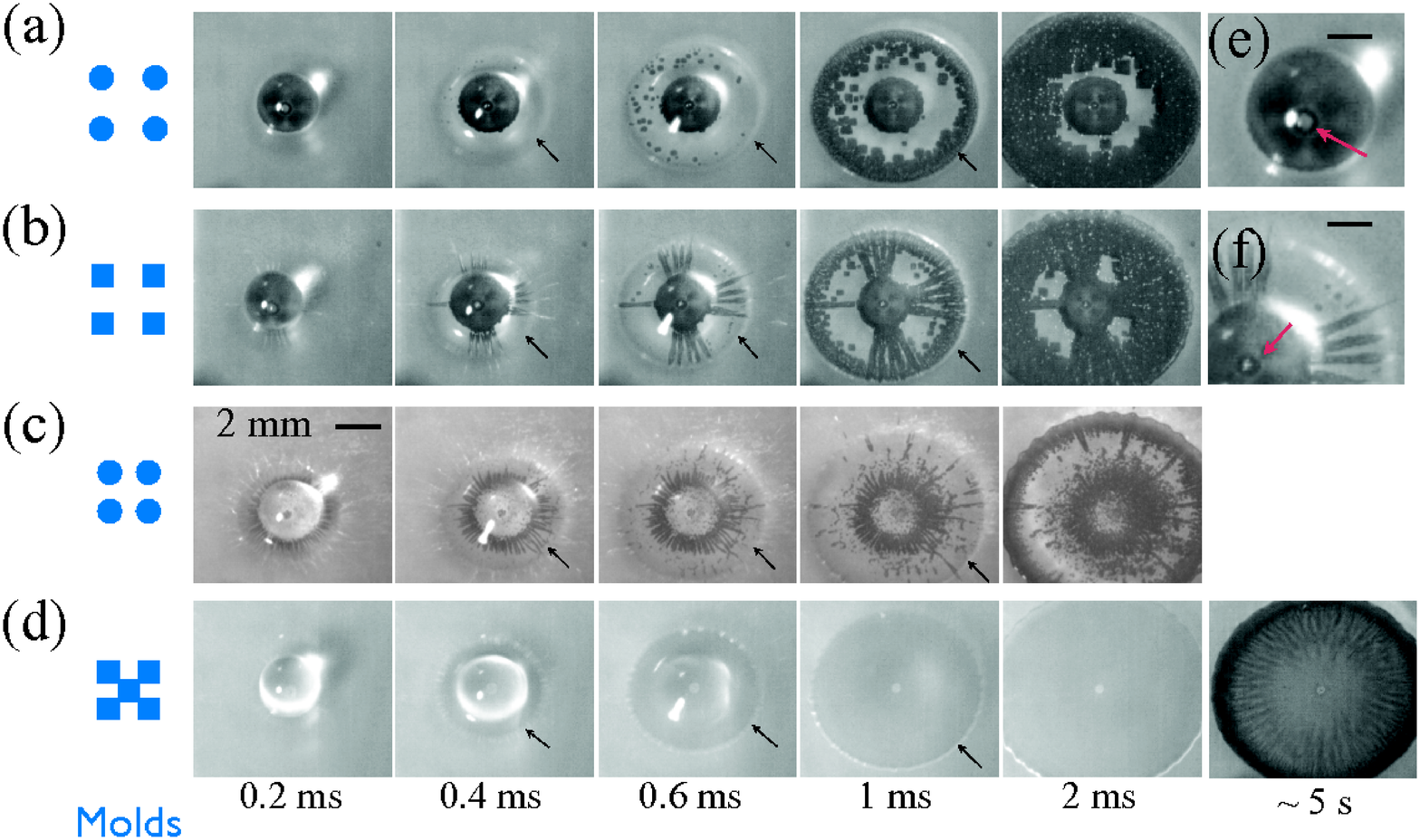}
\caption{\label{top_view_1atm}
(Color online) Top-view evolutions of ethanol drop impact dynamics upon four different micro-patterns (drawn at scale in the left column) at the atmospheric pressure: (a) a deposition, and (b)--(d) splashing. Here $We = 325 \pm 15$ and $R = 1.3$ mm.  The dark areas indicate the liquid wetting regime whereas the bright areas reveal the dry part of the surfaces. The black arrows point out the outer edge of the spreading lamella. In (a) and (b) the droplets wet the central areas at the impact, in (c) the droplet wets a circumference and entraps a large air film at the central zone, and in (d) liquid spreads on top a thin air bubble without wetting the bottom of the surface within $2$ ms. The geometric parameters of the used micro-molds are (a/$\mu m$, b/$\mu m$) = $(5, 9), (5, 7), (1, 4),$ and $(0, 5)$, in the sequence from (a) to (d). (e) and (f) are the close-up snapshots of (a) at $0.2~ms$ and (b) at $0.6~ms$, respectively, showing the tiny, entrapped air packet (marked by the red arrows) under the center of the droplet. The inset bars in (e) and (f) indicate a length scale of $1$ mm. In (b), it becomes particularly evident that the splashing is directional, in the direction of the patterns.
} 
\end{center}
\end{figure*}

The high-resolution, top-view snapshots, complementary to the side-view observations, reveal how liquid microscopically invades the patterned surfaces, and whether an air film is present between the impinging drop and the substrates. The areas wetted by liquid appear dark on the microstructured surface in the top-views; in contrast, the dry spots between the micro-pillars remain transparent. Fig.~\ref{top_view_1atm} shows the top-view evolutions of ethanol drop impact onto slightly different micro-arrangements of pillars at the same $We = 325 \pm 15$. Intriguingly, the central zone at the impact is largely wetted by ethanol droplets for the square latticed micro-pillars with a larger interspace of $a = 5 \mu m$, as shown in (a) for a deposition and in (b) for splashing, whereas the central area remains dry for small $a \lesssim 2~\mu m$ in (c).

From the snapshots, within $1$ ms after the impact in Fig.~\ref{top_view_1atm} (a) and \ref{top_view_1atm}(b), a tiny air bubble of width of $O(200~\mu m)$ can be noticed at the center, as shown in Fig.~\ref{top_view_1atm} (e) and (f), surrounded by a large wetted area of the width of $O(2.5~mm)$. Moreover, in (b) droplets are jetting out (between 0.2 and 0.6 ms) in a fourfold direction, which reflects the square patterned arrangement of the micro-pillars, while the ethanol lamella is spreading out. Such directional splashes have before been observed~\cite{LXu_PRE_2007_1, DQuere_private}. It is interesting to note that the thin lamella front (whose outer edge are marked by the arrows in Fig~\ref{top_view_1atm}) spreads over the top of micro-pillars from the snapshots at 0.4 and 0.6 $ms$. Eventually, the liquid wets the microstructures from the slow propagating, outer edge (see snapshot at $1$ ms in (b)).

The geometric parameters of micro-pillars in (a) deposition and (b) splashing are slightly different: both have interspace $a = 5~\mu m$, but $b = 9~\mu m$ in (a) and $b = 7~\mu m$ in (b). The larger $b$ in (a) allows easy air flow through the porous microstructures. In contrast, for a smaller $a \approx 1.3~\mu m$ in (c), the impacting droplet merely wetted the surface with a contacting perimeter while a large central zone remained dry. In addition, more liquid jets in comparison to (b) are emitted out from the contacting boundary. This reveals the important role of the amount and the drainage of the intervening gas between the impinging droplet and the solid surface in the splash~\cite{MBrenner_PRL_2009,MBrenner_JFM_2010}. In (d), the ethanol droplet spreads over the surface with a chessboard-like pattern with an air film in between for a couple of ms after the impact. The snapshots in (d) remain bright in the early time, and afterwards the ethanol liquid completely wetted the surface, recognized as the darker image shown in the last frame. In the chessboard-like structures, the width of the micro-pillars is about $6.5~\mu m$ and the periodicity $d = 10~\mu m$, analyzed from the SEM images. This closely-packed chessboard-like patterned surface, providing no interconnecting air flow pathways between the micropillars, mimics the drop impact problem onto a flat surface, for which unfortunately the wetting zone and the three-phase contact is difficult to observe.

The top-view recordings reveal the crucial role of the air in the presented outcomes of the splash. Firstly, there is entrapped air: a tiny air bubble in the center is observed in (a) and (b), although large areas around the bubble are wetted; large dry areas appear in (c) and (d) for smaller micro-spacing pillars with no easy interconnecting path for air to drain out. More importantly, how air escapes from and interact with the liquid lamella over the solid surface affect the splashing dynamics, showing distinctive behaviors with different micropatterns as revealed in Fig.~\ref{side_view_1atm}. 
From (a) to (c), the outward spreading liquid starts penetrating and wetting the microstructures at the leading-edge of the lamella after a few ms and thus no further development of fingering structures occur, which may lead to more breakups of droplets in the case of non-wetting liquids. In contrast to Figs.~\ref{top_view_1atm}a, b, c, Fig.~\ref{top_view_1atm}d presents a study mimicking ethanol drop impact onto a flat surface owning to the used chessboard-like micro-mold with which air can not flow between the microstructures.  Indeed, from the high-speed photographs, the bottom of the microstructures in (d) initially stays dry, as shown by the bright images in the early time of a couple ms. The liquid sheet almost horizontally spreads outward over a thin air film, resulting in slightly upward jetting droplets (see Fig.~\ref{side_view_1atm} c).  The present experimental observations reveal the existence of the air films for ethanol splashing impact, but unfortunately we presently can not measure their microscopic thickness. Nevertheless, these top-view observations show that the drainage of the squeezed gas film as the droplet approaches the solid surface can produce a splash beyond the critical $We_c$. In addition, the amount of entrapment of air affects the intensity of splashing; for instance in (b) a few satellite droplets are produced when the central wetted zone is large and in (c) more droplets are jetted out when the continuous air film is present. In short, the entrapped and squeezed gas film, taking place prior to the impact, plays a critical role in splashing impact, as suggested by Refs.~\cite{Chandra_PoF_2003, MBrenner_PRL_2009, MBrenner_JFM_2010} based on calculations of the time evolutions of the gas pressure and the lqiuid-gas interface as the drop impinges a flat surface.
	
\subsection{The influence of air pressure}

\begin{figure}
\begin{center}
\includegraphics[width=3.2in]{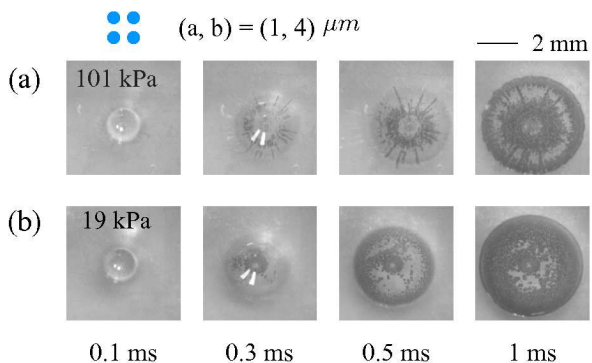}
\caption{\label{top_views_lowP}
(Supplementary movies) Suppression of ethanol splashing by a reduced air pressure. Top-view evolutions of ethanol drop impact onto the same micro-patterned substrate at different air pressure: (a) splashing at atmospheric pressure $P_{air} = 101$ kPa and (b) deposition at a lower pressure $P_{air} = 19$ kPa. The droplet radius was $0.75$ mm, released from the same height for the two cases. Here $We = 190 \pm 10$ in (a) and $We = 225 \pm 10$ in (b). 
}
\end{center}
\end{figure}

\begin{figure}
\begin{center}
\includegraphics[width=3.2in]{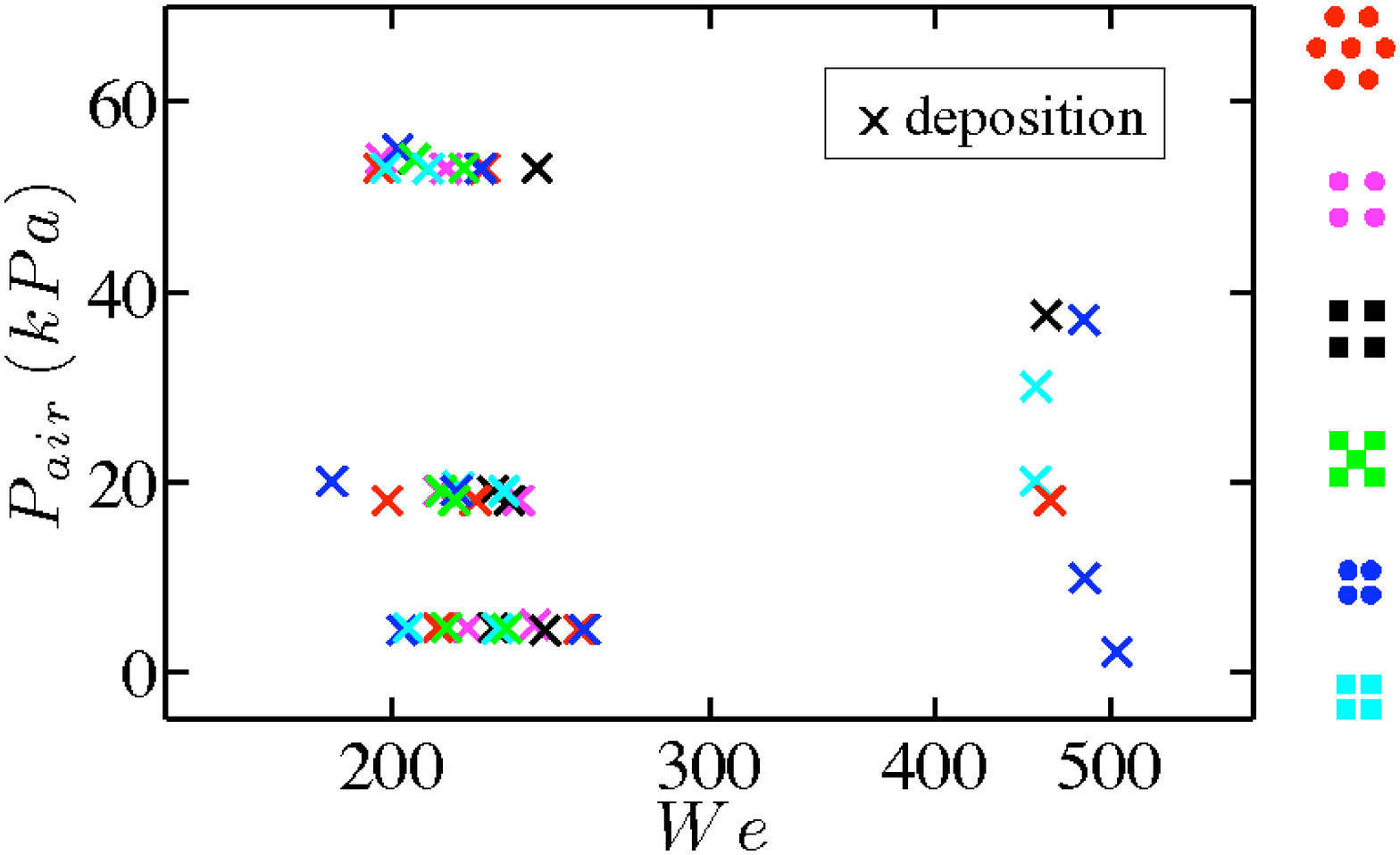}
\caption{\label{phase_diagram_lowP}
(Color) Fraction of the phase diagram of ethanol drop impact at reduced air-pressure, showing liquid deposition onto all the used micro-patterns and thus the suppression of splashing impact. The same replica molds were used as those presented in Fig.~\ref{phase_diagram}.
}
\end{center}
\end{figure}

Next, we control the surrounding air pressure $P_{air}$ to verify the vital effect of air on splashing~\cite{LXu_PRL_2005}.  When $P_{air}$ was decreased, originally splashing ethanol droplet at $P_{air} = 1$ atm now at a reduced air pressure showed a gentle deposition in the explored parameter regime. Fig.~\ref{top_views_lowP} shows the comparison of ethanol droplets impacting onto the indicated microstructures with the same releasing height for different $P_{air}$: we find (a) violent splashing at $P_{air} = 1$ atm, and (b) gentle deposition at $P_{air} = 19$ kPa.  The top-view snapshots reveal a largely dry central area for $P_{air} = 1$ atm in (a), in contrast to the large wetted central region in (b) at $P_{air} = 19$ kPa. Fig.~\ref{phase_diagram_lowP} shows that a deposition occurs for all the used micropatterns at a decreased air pressure in the explored parameter regime. This phase diagram is in vast contrast to that of Fig.~\ref{phase_diagram} for ambient pressures.

\section{Conclusions}

In summary, we present both top-view and side-view, high-speed photographs of ethanol drop impacts onto a variety of microstructues of the height of $h = 5\mu m$ with controlled surrounding air pressure. At small impact velocity a simple spreading of the lamella occurs for all the substrates, whereas at high impact velocity the drop is splashing, ejecting slightly upward lamella from which tiny droplets are emitted within $0.5$ ms, can happen.  The splash threshold expressed in terms of $We_{c}$ for the different micro-patterns was experimentally investigated. ${We}_{c}$ is about twice smaller for the substrates of closely packed micro-pillars of $d \approx 7 \mu m$ than that for dilutely packed ones of $d = 10 \mu m$.  Our top-view observations reveal the entrapment of an air film between the liquid and the used solid surface prior to the splashing, as suggested by a recent theory~\cite{MBrenner_PRL_2009}. These images show the crucial influence of the entrapped air film which can alter the splashing dynamics. In the investigated parameter regime, for all the micropatters the splash can be eliminated through reducing the air pressure. The top-view snapshots under decreased air pressure show a small dry central area in comparison to those at atmospheric pressure. As a result, we think that the drainage of the squeezed entrapped air is responsible for the splashing observed in the ethanol drop impact onto the microstructures. This reveals the importance of the surface micro-structures, allowing for or blocking pathways for air flow. In this way even directional splash can be achieved. 

{\bf Acknowledgment} The authors gratefully thank A. Prosperitti and C. Pirat for the stimulating and useful discussion, S. Pacheco for the SEM pictures, and the Membrance Technology Group at the University of Twente for the micro-patterned molds.

\bibliography{Tsai_ethanol_splashing_impact_submitted_arxiv}

\end{document}